\documentstyle[preprint,aps]{revtex}

\begin{document}
\draft
\preprint{ IFUM 505/FT\\ 
\quad CERN-TH/95-163\\  
\quad hep-ph/9505432}
\title{Perturbative evaluation\\ of the eigenvalues of the Herbst Hamiltonian}
\author{Nora Brambilla$^*$ and Antonio Vairo$^{**}$}
\address{
$^*$Dipartimento di Fisica dell'Universit\`{a} -- Milano\\
INFN, Sezione di Milano, Via Celoria 16, 20133 Milano, Italy;\\
$^{**}$ CERN, Theory Division, CERN, CH-1211 Geneva 23, Switzerland,\\
INFN, Sezione di Bologna, Via Irnerio 46, 40126 Bologna, Italy}
\maketitle
\begin{abstract}
We reconsider the well-known and long-debated problem of the 
calculation of the eigenvalues of the Herbst Hamiltonian
$2\sqrt{ p^2 +m^2} - \kappa / r$. We give a formulation 
of the problem that allows, for the first time, a perturbative evaluation
of the eigenvalues for any $n$ and $l$, and in principle 
up to any order in $\kappa$ via standard Kato perturbation 
theory. We present the evaluation of the  energy of the $n=1$ and 
$n=2$ states up to $\kappa^6$, confirming the result previously obtained by 
Le Yaouanc et al. with a completely different technique. 
Moreover we give the $n=2, l=1$ level, which is new. Discussion of the results 
and comparison with previous findings are given at the end.
\end{abstract}
\vspace{0.5in}
\par CERN-TH/95-163 
\par June 1995


\newpage
\section{INTRODUCTION}
In the last years the so-called Herbst Hamiltonian (sometimes also called 
the spinless relativistic Coulomb Hamiltonian): 
\begin{equation}
H= 2 \sqrt{{\bf p}^2 + m^2} -{\kappa\over r},
\label{eq:herbeq}
\end{equation}
has been intensively studied in terms of both the spectrum and 
the eigenfunctions \cite{herb,dur,cast,such,luch,oliv}.  
The reason of the interest in such a Hamiltonian is not only a mathematical one 
\cite{herb}  but mainly a physical one, also due to the relevance 
of (\ref{eq:herbeq}) in  quarkonium phenomenology \cite{mot}.
In fact we recall that, for an appropriate potential $V$ the Hamiltonian 
$H= 2 \sqrt{{\bf p}^2 + m^2} + V(\bf{x})$ is the spinless Salpeter equation
representing a well-defined standard approximation to the Bethe--Salpeter 
formalism  for the description  of quark-antiquark bound states
\cite{luch,luchrep,salpvar}; in this context Eq. (\ref{eq:herbeq}) 
is suitable for the treatment of short-range effects in bound $q \bar{q}$ 
systems and for the study of decay rates; moreover it was also applied 
to the study of boson stars \cite{mar}.
\par
For the reasons presented above, a lot of attention has been addressed 
in the last years to the calculation of the ground state $E_{10}$ of the Herbst 
Hamiltonian. First, an exact closed expression was found in Ref.
\cite{dur} which,  however, turned out to be wrong \cite{such,dur}; then, 
many analytic estimates of lower \cite{herb,mar} and upper \cite{luch} bounds 
were given for $E_{10}$. Up to now we are aware of only a paper devoted to the 
perturbative calculation of $E_{10}$ \cite{oliv}. 
The method used in \cite{oliv} has the merit of being systematic and 
allows the authors to settle an iteration procedure on a reasonably rigorous 
basis
\footnote{The authors of Ref.\cite{oliv} derive an equation for a 
function related to the eigenfunction and use it to obtain contributions up 
to $\kappa^5$ to the $l=0$ levels. Then, they use the analyticity properties 
of the wave function to obtain next to leading orders up to 
$\kappa^7$.}, 
but it is not general.  Its starting point is a peculiar representation of 
the Hilbert space of $l=0$ states (introduced in \cite{dur}) in which the 
Coulomb potential operator has a particularly simple form.
\par
For this reason we have decided that it could be useful and instructive to 
apply a general and powerful perturbative technique to the 
calculation of the eigenvalues of the Herbst Hamiltonian. 
This method, not being  founded on peculiar transformations, can in principle 
be applied for any value of the angular momentum and also extended to the 
consideration of potentials different from the Coulombian one. 
In fact it was proposed for the first time for the calculation of the
positronium energy levels in Ref. \cite{remiddi}. 
\par 
The plan of the paper is as follows. In Sec. II we explain the 
perturbative technique; in Sec. III we apply it to the evaluation 
of the energy levels in the case $n=1$ and  $n=2$ ($l=0,1$); 
in Sec. IV we make a comparison with the results previously obtained in the 
literature and  draw some conclusions.

\section{DESCRIPTION OF THE METHOD}

We have to solve the equation
\begin{equation}
H \Phi = E \Phi
\label{eq:inham}
\end{equation}
with the two-body relativistic Coulombian Hamiltonian given in 
(\ref{eq:herbeq}) or equivalently with the one-body Hamiltonian
\begin{equation}
H= \sqrt{ {\bf p}^2 + m^2 } -{\alpha \over r}.
\label{one}
\end{equation}
The energy levels of the two-body (\ref{eq:herbeq}) and  the one-body 
(\ref{one}) cases may easily be related by identifying  both mass and Coulomb 
strength coupling parameters according to $ m \rightarrow 2 m $ and 
$\alpha \rightarrow {\kappa / 2}$. In the following we consider 
the one-body case.
\par
The perturbative  solution of Eq. (\ref{eq:inham}) which comes immediately 
to mind refers to the well-known non-relativistic Schr\"odinger Hamiltonian 
as the unperturbed starting point and  supplies the eigenvalues $E_{nl}$ 
$(n=1,2 \dots; l=0,1, \dots ,n-1  )$ by evaluating perturbatively the 
expectation value of the relativistic corrections, coming from the 
expansion of the kinetic square root, on the analytically available
zeroth-order eigenfunctions. This can be simply done for the contribution in 
the fourth power of ${\bf p}$. For instance, for the  ground state 
zeroth-order energy $E_{10}$, we can calculate
\begin{equation}
\delta E_{10} = \langle \varphi_{100} \vert - {{\bf p}^4 \over 
 8 m^3 } \vert \varphi_{100}\rangle 
= - {5 \over 64 } m \alpha^4.
\end{equation}
where $\varphi_{100}$ is the ground state Schr\"odinger--Coulomb 
wave-function:
\begin{equation}
\varphi_{100} ({\bf p})= { 8 \sqrt{\pi} (m\alpha)^{5\over 2} \over
( {\bf p}^2 + (m\alpha)^2 )^2} \,    . 
\end{equation}
However, already with the sixth power of $ {\bf p}$, 
one encounters a divergent integral; in order to get the next term in 
the perturbative expansion, which turns out to be of order $\alpha^5$, one 
has to sum up an infinite number of contributions coming from 
all the possible intermediate states (cf. Ref.\cite{oliv}).
In general, the relativistic kinetic energy $\sqrt{{\bf p}^2 +m^2}$,
behaving as $ \vert {\bf p}\vert $ for very large $\vert {\bf p}\vert$,
cannot be approximated by any finite polynomial in ${\bf p}^2$.
This seems to prevent the possibility of profiting from the known results 
for the Schr\"odinger equation in the relativistic case. 
\footnote{
At least, not without some ad hoc transformations and intricate 
calculations as in Ref.\cite{oliv}
}
There is, however, a trick for recovering, at least formally, a ``kinetic 
energy'' ${\bf p}^2 $, even starting with $ \sqrt{{\bf p}^2 + m^2}$; 
the price to pay is a kind of energy-dependent potential instead of the 
Coulomb one \cite{remiddi}. Let us consider Eq. (\ref{eq:inham}) 
written in the momentum space
\begin{eqnarray}
& & \int {d^3 {\bf q} \over ( 2 \pi )^3 } 
\left[ ( 2 \pi )^3 \delta^3 ({\bf p}- {\bf q})
( \sqrt{ {\bf p^2}+m^2} - E) + V({\bf p} - {\bf q}) \right] \Phi({\bf q})=0.
\label{eq:momham}\\
& & V({\bf p}) = -{4\pi\alpha \over {\bf p}^2}.
\nonumber
\end{eqnarray}
We define 
\begin{equation}
R(E;{\bf p}) \equiv \left[ {2 m \over E_p + E} \right]^{1\over 2} \,   ; 
\quad \quad E_p= \sqrt{{\bf p}^2 +m^2}
\label{eq:defr}
\end{equation}
and put
\begin{eqnarray}
\Phi ( {\bf p}) & =& R^{-1} (E;{\bf p}) \phi({\bf p})\nonumber \\
V({\bf p}-{\bf q}) & =& R(E;{\bf p}) v(E;{\bf p}, {\bf q}) R(E;{\bf q}).
\label{eq:multdef}
\end{eqnarray}
By using Eqs. (\ref{eq:defr}) and (\ref{eq:multdef}), we rewrite 
Eq. (\ref{eq:momham}) in the form
\begin{equation}
\int {d^3 {\bf q}\over (2 \pi )^3 } 
\left[ ( 2 \pi )^3 \delta^3 ({\bf p}-{\bf q}) 
\left( {{\bf p}^2\over 2 m} - { E^2 - m^2\over 2 m} \right) 
+ v(E;{\bf p}, {\bf q})\right]
\phi ({\bf q}) =0
\label{eq:schmod}
\end{equation}
which is  formally of the Schr\"odinger type, i.e. with a quadratic
kinetic term, {\it but} with an energy-dependent interaction. 
Equation (\ref{eq:schmod}) is the appropriate starting point for the 
application of a perturbative technique.
\par 
Let us rephrase the same transformation in the propagator formalism, which
is more appropriate  to treat an energy-dependent perturbation term.
With reference to Eq. (\ref{eq:momham}) we can write for the one-particle 
propagator $G$ :
\begin{equation}
G({E;\bf p}, {\bf q}) = G_0(E;{\bf p}) 
\left[ (2 \pi )^3 \delta^3 ({\bf p}- {\bf q}) 
+ \int { d^3 {\bf k} \over (2 \pi )^3} V({\bf p}- {\bf k})
 G({E;\bf k},{\bf q})\right],
\label{eq:GGG}
\end{equation}
$G_0$ being the free propagator
\begin{equation}
G_0(E; {\bf p}) = {1\over E- E_p}.
\end{equation}
Now, by using the identity
\begin{equation}
{1\over E-E_p } = {1 \over (E^2-m^2)/ 2 m -  {\bf p}^2 / 2 m}
\left( { E+E_p \over 2 m} \right),
\end{equation}
Eqs. (\ref{eq:multdef}), and defining 
\begin{equation}
g_0 (E^*) = {1\over E^* - {{\bf p}^2 / 2 m} }\,  , \quad \quad
 E^*= {E^2 -m^2 \over 2 m}, 
\label{Estar}
\end{equation}
we can rewrite Eq. (\ref{eq:GGG}) as 
\begin{equation}
g(E^*;{\bf p}, {\bf q}) = g_0(E^*;{\bf p}) 
\left[(2 \pi)^3 \delta^3({\bf p}-{\bf q})
+ \int {d^3 {\bf k} \over (2 \pi )^3 } 
v(E;{\bf p},{\bf q}) g(E^*;{\bf k}, {\bf q}),
\right]
\label{eq:ggstar}
\end{equation}
where $v$ is given in (\ref{eq:multdef}) and 
\begin{equation}
g(E^* ;{\bf p}, {\bf q})  = R(E;{\bf p})G(E;{\bf p}, {\bf q}) R(E;{\bf q}).
\end{equation}
We emphasize that in Eq. (\ref{eq:ggstar}) the propagators $g$ and $g_0$  
has to be understood as depending on the modified energy $E^*$, which is a 
function of $E$.
\par
Inserting $V$ in place of $ v$ in (\ref{eq:ggstar}) we obtain the exactly
solvable equation for the well-known Schr\"odinger-Coulomb propagator 
$ g_s$ :
\begin{equation}
g_s(E^* ;{\bf p}, {\bf q} ) = g_0(E^*;{\bf p}) 
\left[(2 \pi)^3 \delta^3({\bf p}-{\bf q})
+ \int {d^3 {\bf k}\over (2\pi)^3 } V({\bf p}-{\bf k}) 
g_s(E^* ;{\bf k},{\bf q})\right].
\end{equation}
The $g_s$ propagator has poles in correspondence of the energy values 
\begin{equation}
E^* = -{m \alpha^2 \over 2 n^2} \quad 
\Rightarrow E = E_n^0 \equiv \sqrt{m^2 -\gamma_n^2}, 
\quad \quad \gamma_n \equiv {m \alpha \over n}, \quad n = 1,2, \dots \quad ,
\label{gamman}
\end{equation}
and can be written for any $n$ as
\begin{eqnarray}
g_s(E^*;{\bf p},{\bf q}) &=& 
{\sum_{l,m}\varphi_{nlm}({\bf p})\varphi^*_{nlm}({\bf q})
\over E^* + m\alpha^2/ 2 n^2} 
+ \hat{g}_n(E^*; {\bf p}, {\bf q}),
\nonumber\\
&=&
  {m\over E_n^0}
{\sum_{l,m}\varphi_{nlm}({\bf p})\varphi^*_{nlm}({\bf q})\over E-E_n^0}
- {m\over E_n^0}
{\sum_{l,m}\varphi_{nlm}({\bf p})\varphi^*_{nlm}({\bf q})\over E+E_n^0}
+ \hat{g}_n(E^*; {\bf p}, {\bf q})
\nonumber\\
&\equiv& 
{m\over E_n^0}
{\sum_{l,m}\varphi_{nlm}({\bf p})\varphi^*_{nlm}({\bf q})\over E-E_n^0}
+ \hat{g}^\prime_n(E; {\bf p},{\bf q}),
\label{gsstar}
\end{eqnarray}
$\varphi_{nlm}$ being the Schr\"odinger--Coulomb wave-functions and
$\hat{g}_n$ (or $\hat{g}^\prime_n$) being  the regular part of the propagator. 
\par
Now, by using standard Kato perturbation theory \cite{kato}, one obtains the 
following expansion for the energy levels ($\delta V \equiv v-V$),
\begin{eqnarray}
E_{nl} &=& E_{n}^0 +  {1\over 2l +1}
\langle \delta V(E_n^0)\rangle_{nl} 
+ \nonumber \\
&+& {1\over 2l+1} \left\{\langle \delta V(E_n^0) \hat{g}_n^\prime
(E_n^0) \delta V(E_n^0) \rangle_{nl} 
+ {1\over 2l+1} \langle \delta V(E_n^0) \rangle_{nl} 
\left\langle {\partial \over \partial E} \delta V (E_n^0) 
\right\rangle_{nl}\right\} 
+ \nonumber \\
&+&  {1\over 2l+1}
\langle \delta V(E_n^0) \hat{g}_n^\prime(E_n^0) 
\delta V(E_n^0) \hat{g}_n^\prime(E_n^0) \delta V(E_n^0) \rangle_{nl} 
+ \dots \quad,
\label{energy}
\end{eqnarray}
where the symbol $\langle \, \, \rangle_{nl}$  stands for 
\begin{equation}
\langle \, \, \rangle_{nl} \equiv 
{m \over E^0_n}\sum_m
\int {d^3 {\bf p} \over ( 2 \pi )^3 } 
\int {d^3 {\bf q} \over ( 2 \pi )^3 } \Big( \quad \Big) 
\varphi_{nlm}({\bf p})\varphi^*_{nlm}({\bf q}) .
\end{equation}
The summation over the quantic number $m$ and the factors 
$1 / (2l + 1)$ takes in account the degeneration of the level.
\par
Expansion (\ref{energy}) is valid for any $n$ and any $l$. 
As higher order in $\delta V$ supply higher-order leading contributions 
in $\alpha$, Eq. (\ref{energy}) is perturbative not only 
in $\delta V$ but also in $\alpha$. So up to a given order in $\alpha$ only 
a finite number of terms in (\ref{energy}) contribute to it.
\par
In the next section we apply Eq. (\ref{energy}) to the evaluation of $E_{nl}$ 
in the cases $n=1$ and $n=2$  $l=0,1$ up to the order $\alpha^6$.

\section{CALCULATION OF THE ENERGY LEVELS}
 
Up to order $\alpha^6$, only the terms explicitly written in Eq. (\ref{energy}) 
have to be taken in to account. It is useful to write ${\hat g}^\prime(E^0_n)$ 
in the form:
\begin{eqnarray}
{\hat g}^\prime(E^0_n) &=& 
{\hat g}\left({-\gamma_n^2\over 2 m} \right) 
- {m\over 2 {E_n^0}^2} \sum_{l,m}\varphi_{nlm}\varphi^*_{nlm} = 
\nonumber\\
&=& g_0\left({-\gamma_n^2\over 2 m}\right) 
+ g_0\left({-\gamma_n^2\over 2 m}\right)\, V \, 
g_0\left({-\gamma_n^2\over 2 m}\right) 
+ \hat{R}^\prime_n(E_n^0) ,
\label{Rcap}
\end{eqnarray}
where 
$$ 
\hat{R}^\prime_n(E_n^0)  \equiv \hat{R}_n\left({-\gamma_n^2\over 2 m}\right)  
- {m\over 2 {E_n^0}^2} \sum_{l,m}\varphi_{nlm}\varphi^*_{nlm} .
$$
The functions $R_1\left(-\gamma_1^2/ 2 m\right)$ and 
$R_2\left(-\gamma_2^2/ 2 m\right)$ are given in the appendix.
From Eq. (\ref{Rcap}) and  
\begin{eqnarray}
&\ & \langle \delta V(E_n^0) \hat{g}_n^\prime(E_n^0) 
\delta V(E_n^0) \hat{g}_n^\prime(E_n^0) \delta V(E_n^0) \rangle_{nl} 
= \nonumber\\
&\ & \quad \quad \quad \quad 
= \left\langle \delta V(E_n^0)\, g_0\left({-\gamma_n^2\over 2 m}\right)\,  
\delta V(E_n^0)\,  g_0\left({-\gamma_n^2\over 2 m}\right)  \, 
\delta V(E_n^0) \right\rangle_{nl}  + o(\alpha^6) ,
\nonumber
\end{eqnarray}
we obtain the only relevant contributions to the energy levels 
up to order $\alpha^6$ :
\begin{eqnarray}
E_{nl} &=& E_{n}^0 +  {1\over 2l +1}
\langle \delta V(E_n^0)\rangle_{nl} 
+ {1\over 2l+1} 
\Bigg\{
\left\langle \delta V(E_n^0)\,  g_0\left({-\gamma_n^2\over 2 m}\right) \, 
\delta V(E_n^0) \right\rangle_{nl} + 
\nonumber\\
&\ & \quad \quad \quad  + \left\langle \delta V(E_n^0) \, 
g_0\left({-\gamma_n^2\over 2 m}\right)\,  V 
\, g_0\left({-\gamma_n^2\over 2 m}\right) \, 
\delta V(E_n^0) \right\rangle_{nl} 
+ \langle \delta V(E_n^0) \hat{R}^\prime_n(E_n^0) \delta V(E_n^0) \rangle_{nl} +
\nonumber\\
&\ & \quad \quad \quad  + {1\over 2l+1} \langle \delta V(E_n^0) \rangle_{nl} 
\left\langle {\partial \over \partial E} \delta V (E_n^0) 
\right\rangle_{nl}\Bigg\} 
+ \nonumber \\
&+&  {1\over 2l+1}
\left\langle \delta V(E_n^0) \, g_0\left({-\gamma_n^2\over 2 m}\right) \, 
\delta V(E_n^0) \, g_0\left({-\gamma_n^2\over 2 m}\right) \, 
\delta V(E_n^0) \right\rangle_{nl} 
+ o(\alpha^6).
\label{energy6}
\end{eqnarray}
With the symbol $o(\alpha^6)$ we indicate higher-order contributions, 
typically starting with $\alpha^7$ and $\alpha^7 \ln  \alpha$ terms.
\par
First let us consider the case $l=0$. Using the Schr\"odinger equation 
($ g_0 V \varphi = \varphi $) we can perform  some straightforward 
cancellations and reduce Eq. (\ref{energy6}) to the form:
\begin{eqnarray}
E_{n0} &=& E_n^0 
+ \left\langle v(E_n^0)\,g_0\left({-\gamma_n^2\over 2 m}\right)\,  
v(E_n^0) \, g_0\left({-\gamma_n^2\over 2 m}\right)\,  
\delta V(E_n^0)\right\rangle_{n0} 
+ \left\langle {\partial \over \partial E} \delta V (E_n^0)\right\rangle_{n0} 
\langle \delta V (E_n^0) \rangle_{n0}
\nonumber\\
&+& 
\langle \delta V(E_n^0)\hat{R}^{\prime}_n(E_n^0)\delta V(E_n^0)\rangle_{n0} 
+ o(\alpha^6) .
\label{energy60}
\end{eqnarray}
For the level $n=1$ we obtain:
\begin{eqnarray}
{\langle v g_0 v g_0 \delta V \rangle_{10} \over m} &=&  
-{1\over 2} \alpha^4 + {8\over 3} {\alpha^5 \over \pi} 
- \alpha^6 \ln \alpha^{-1} + 
\nonumber \\
&+& \left({1\over 8} - {1\over 4}\zeta(3) 
- {2\over \pi^2} + {7 \over \pi^2}\zeta(3)
- {\pi^2 \over 8}\right) \alpha^6 + o(\alpha^6),
\nonumber \\
{\langle \delta V \rangle_{10} 
\langle {\partial \over \partial E}\delta V \rangle_{10}\over m} 
&=& {1\over 4} \alpha^6 + o(\alpha^6),
\nonumber\\
{\langle \delta V \hat{R}^\prime_1 \delta V\rangle_{10}\over m}
&=& \left( {\pi^2 \over 8 }
 - {19\over 8} + {1\over 4}\zeta(3) \right) \alpha^6 + o(\alpha^6).
\nonumber
\end{eqnarray}
Inserting the various contributions in Eq. (\ref{energy60}), 
we obtain the final result:
\begin{equation}
{E_{10}\over m} = 
1- { 1\over 2} \alpha^2 - {5 \over 8} \alpha^4 
+ {8 \over 3}{\alpha^5 \over \pi} 
- \alpha^6 \ln \alpha^{-1} + \left({7 \over \pi^2 } \zeta(3) 
- {2\over \pi^2 } - {33\over 16}\right) \alpha^6 + o(\alpha^6).
\label{e10}
\end{equation}
Similarly for the $n=2, l=0$ case we calculate 
\begin{eqnarray}
{\langle v g_0 v g_0 \delta V\rangle_{20}\over m} &=& 
-{3\over 32} \alpha^4
+{1\over 3} {\alpha^5\over\pi}  
- {1\over 8} \alpha^6 \ln \alpha^{-1}+ 
\nonumber \\
&+& \left( {127\over 512} -{1\over 8} \zeta(3) 
-{1\over 4}{1\over\pi^2} + {7 \over 8}{1\over \pi^2}\zeta(3) 
-{\pi^2 \over 32} 
-{1\over 8}\ln 2 \right)\alpha^6 
+ o(\alpha^6),
\nonumber\\
{\langle \delta V \rangle_{20}
\langle {\partial \over \partial E}\delta V \rangle_{20}\over m} 
&=& {3\over 256} \alpha^6 + o(\alpha^6),
\nonumber\\
{\langle \delta V \hat{R}^\prime_2 \delta V\rangle_{20}\over m}
&=& \left( {1\over 32}\pi^2  - {231\over 512} 
+ {1\over 8} \zeta(3) \right) \alpha^6 + o(\alpha^6),
\nonumber
\end{eqnarray}
and then obtain
\begin{eqnarray}
{E_{20}\over m} &=& 
1- { 1\over 8} \alpha^2 - {13 \over 128} \alpha^4 + {1 \over 3}
{\alpha^5 \over \pi } - {1\over 8} \alpha^6 \ln \alpha^{-1} + 
\nonumber\\
&\ & \quad \quad \quad 
+ \left({7 \over 8}{1\over \pi^2}\zeta(3)  - {1\over 4}{1\over \pi^2 }  
- {197\over 1024} -{1\over 8}\ln 2 \right) \alpha^6 + o(\alpha^6).
\label{e20}
\end{eqnarray} 
\par
When  $l$ is different from zero, we have to apply directly 
Eq. (\ref{energy6}). Let us consider the case $n=2, l=1$. Then 
\begin{eqnarray}
{\langle  \delta V \rangle_{21}\over m} &=& 
-{1\over 32}\alpha^4 + {19 \over 512}\alpha^6 + o(\alpha^6),
\nonumber\\
{\langle  \delta V g_0 \delta V \rangle_{21}\over m} &=& 
\left( {1 \over 96}\pi^2 -{15\over 128} \right) \alpha^6 + o(\alpha^6),
\nonumber\\
{\langle  \delta V g_0 V g_0 \delta V \rangle_{21}\over m} &=& 
\left( {1 \over 48}\pi^2 + {1\over 8}\zeta(3) 
- {93\over 256} \right) \alpha^6 + o(\alpha^6),
\nonumber\\
{\langle \delta V \hat{R}^\prime_2 \delta V\rangle_{21}\over m} &=&
\left( {2051\over 4608} - {1\over 8}\zeta(3) 
- {1 \over 32}\pi^2 \right) \alpha^6 + o(\alpha^6).
\nonumber\\
{\langle \delta V \rangle_{21}
\langle {\partial \over \partial E}\delta V \rangle_{21}\over m} 
&=& {3\over 256} \alpha^6 + o(\alpha^6),
\nonumber\\
{\langle  \delta V g_0 \delta V g_0 \delta V \rangle_{21}\over m} &=& 
o(\alpha^6).
\nonumber
\end{eqnarray}
and, summing up all the contributions, we obtain
\begin{equation}
{E_{21}\over m} = 
1- { 1\over 8} \alpha^2 - {7 \over 384}\alpha^4 + {25\over 27648}\alpha^6
+ o(\alpha^6),
\label{e21}
\end{equation} 
which, surprisingly, does not contain contributions in $\alpha^5 $ 
and $ \alpha^6 \ln {\alpha}$.
\par
The description (in another context) of some techniques used to extract 
 from the integrals the  various order in powers of $\alpha$ can be found 
in \cite{vairo}. All calculations were performed with the help 
of the program of symbolic manipulations FORM \cite{ver}.

\section{DISCUSSION AND CONCLUSIONS}

In conclusion, in the case $n=1,2$, $l=0$, we have obtained a result 
(Eqs. (\ref{e10}) and (\ref{e20})) that confirms, up to order 
$\alpha^6$, the previous findings of Le Yaouanc et al. \cite{oliv}. 
However, the result obtained in the case $n=2, l=1$ (Eq. (\ref{e21}))
is new and displays also a different analytical structure.
\par
We emphasize again that our method is general and can be applied 
to the calculation of the eigenvalue of the Herbst Hamiltonian
for any $n$ and any $l$, and in principle up to the desired order 
in $\alpha$. Moreover, due to its generality, this method could be applied 
also in the case of a potential different from the Coulomb one.
We notice that the problem can be solved on a general ground and by well-known 
energy-dependent perturbation theory due to the appropriate formulation 
of the starting point in (\ref{eq:schmod}). In this way, in fact, we gain an 
energy-dependent perturbation containing $\alpha$.
\par
Finally let us mention some other points:
\begin{description}
\item{1)} The result we have obtained for the ground-state energy fits nicely 
in the upper and lower  variational bounds given in  Refs. \cite{luch,mar}.
\item{2)} The perturbative expansion of the eigenvalues given in 
(\ref{e10})--(\ref{e21}) 
is suitable to obtain good numerical estimates of the energy of the level 
in principle 
up to the critical value of the coupling constant $ \alpha= { 2 / \pi}$.
As an example, we present in Table 1 the evaluation of $E_{10}$, $E_{20}$ 
and $E_{21}$ for different values of $ \alpha$. We have chosen the same 
values of $\alpha$ as in 
Ref. \cite{mar}, so that it could be appreciated (for small couplings at 
least) the improvement introduced by this perturbative expansion 
with respect to the variational evaluation (cf. the variational upper 
$E_{10}^{\max}$ and lower $E_{10}^{\rm min}$ estimates  given in 
Table 2 of Ref.\cite{mar} and transcribed here in columns 2 and 3 of Table 1).
In Table 1, the last figure in  our results has been put in parenthesis 
because it is affected by the error coming from the neglected $\alpha^7$ 
and $ \alpha^7 \ln{\alpha}$  contributions.
For $\alpha > 0.5$ the error obviously  grows and we have therefore not 
considered significant to present the results and to make a comparison 
also in this case. 
\end{description}

\vspace{0.5in}
\par\noindent
{\bf Acknowledgements}
\vspace{0.2in}
\par\noindent
We would like to thank Wolfgang Lucha for useful discussion and encouragement, 
and Stefano Laporta for the evaluation of some integrals.
We are extremely grateful to Andr\'e Martin for pointing to our attention an error 
in the final result for the $n=2$, $l=1$ state.

\appendix

\section{}

The non-singular part ${\hat g}_n$ of the Schr\"odinger--Coulomb propagator 
is obtained by expanding the analytical expression of $g_s$ in the vicinity 
of the energy level and by subtracting the pole contribution. The expression 
of ${\hat g}_1$ evaluated in $-\gamma_1^2/2 m$ reads (cf. \cite{reg}):
\begin{eqnarray}
\hat{g}_1\left({-\gamma_1^2\over 2 m};{\bf p},{\bf q}\right) &=& 
{(2 \pi )^3 \delta^3 ({\bf p} - {\bf q} ) \over  
-\gamma_1^2/2 m - {\bf p}^2 / 2 m }
+ {1\over -\gamma_1^2/2 m - {\bf p}^2 / 2 m} V({\bf p}- {\bf q} ) 
  {1\over -\gamma_1^2/2 m - {\bf q}^2 / 2 m} - 
\nonumber\\
&-& {64 \pi m \gamma_1^3 \over ({\bf p}^2 + \gamma_1^2 )^2 ( {\bf q}^2
+ \gamma_1^2 )^2 } \left( {5\over 2} 
- 4{\gamma_1^2 \over {\bf p}^2 + \gamma_1^2} 
- 4{\gamma_1^2 \over {\bf q}^2 + \gamma_1^2} \right) - 
\nonumber \\
&-& 
{64 \pi m \gamma_1^3 \over ({\bf p}^2 + \gamma_1^2 )^2 
({\bf q}^2+ \gamma_1^2 )^2} 
\left\{ {1\over 2} \ln C_1 + { 2 C_1 -1 \over \sqrt{4 C_1-1} }
{\rm arctg } \sqrt{ 4 C_1 -1} \right\}
\nonumber \\
& \equiv & g_0 \left( -{\gamma_1 ^2 \over 2 m} \right) 
+ g_0\left( -{\gamma_1 ^2 \over 2 m} \right)\,  V \,
  g_0\left( -{\gamma_1 ^2 \over 2 m} \right)  
+ \hat{R}_1\left( -{\gamma_1 ^2 \over 2 m} \right) .
\end{eqnarray}
In a similar way one obtains the regular part corresponding to the 
$n=2$ pole, which evaluated in $-\gamma_2^2/2 m$ reads:
\begin{eqnarray}
\hat{g}_2\left({-\gamma_2^2\over 2 m};{\bf p},{\bf q}\right) &=& 
{(2 \pi )^3 \delta^3 ({\bf p} - {\bf q} ) \over  
-\gamma_2^2/2 m - {\bf p}^2 / 2 m }
+ {1\over -\gamma_2^2/2 m - {\bf p}^2 / 2 m} V({\bf p}- {\bf q} ) 
  {1\over -\gamma_2^2/2 m - {\bf q}^2 / 2 m} -  
\nonumber\\
&-& 
{256 \pi m \gamma_2^3 \over ( {\bf p}^2 + \gamma_2^2)^2 
({\bf q}^2 + \gamma_2^2 )^2} 
\Bigg\{{2 \gamma_2^2 ( {\bf p} - {\bf q} )^2 \over 
({\bf p}^2 + \gamma_2^2) ( {\bf q}^2 + \gamma_2^2 )}
\left( -{9\over 2} + { 6 \gamma_2^2 \over { \bf p}^2 + \gamma_2^2 }
+ {6 \gamma_2^2 \over {\bf q}^2 + \gamma_2^2 } \right) + 
\nonumber\\
&\,&\quad \quad \quad \quad \quad
+ {3 \over 2} - { 4 \gamma_2^2 \over {\bf p}^2 + \gamma_2^2 } 
- { 4 \gamma_2^2\over {\bf q}^2 + \gamma_2^2 } - 
\nonumber \\
&\,&\quad \quad \quad \quad \quad
- \left( {1 \over 2 C_2} - 1\right) \ln C_2 + 
{ 2 C_2 -4 + 1 / C_2 \over \sqrt{4 C_2 -1}}\,  
{\rm arctg} \sqrt{4 C_2-1} \Bigg\} 
\nonumber \\ 
& \equiv & g_0 \left( -{\gamma_2 ^2 \over 2 m} \right) 
+ g_0\left( -{\gamma_2 ^2 \over 2 m} \right)\,  V \,
  g_0\left( -{\gamma_2 ^2 \over 2 m} \right)  
+ \hat{R}_2\left( -{\gamma_2 ^2 \over 2 m} \right) , 
\end{eqnarray}
with
$$
C_n = { ( {\bf p}^2 + \gamma_n^2 ) ( {\bf q}^2 + \gamma_n^2 ) 
\over 4 \gamma_n^2 ( {\bf p} - {\bf q} ) ^2 },
$$ 
and $\gamma_n$ given in Eq. (\ref{gamman}).

\newpage

\begin{table}
\vspace{2.5in}
\caption{Calculation of the energy levels given in Eqs. (24),(25),(26).
 In columns 2 and 3 are shown the minimum $E_{10}^{\rm min}$
 and the maximum $E_{10}^{\rm max}$ variational estimates of 
 the ground-state energy as given in Table 2 of [10] are shown. 
 The figure in  parenthesis is of the same order as the uncalculated 
 $\alpha^7$ and $\alpha^7 \ln \alpha$ contributions.}
\vspace{0.5in}
\begin{tabular}{|l|l|l|l|l|l|}
$\alpha$       
&$E_{10}^{\rm min}/m$[10]&$E_{10}^{\rm max}/m$[10]&
$E_{10}/m$\quad \quad \quad \quad \quad \quad &
$E_{20}/m$\quad \quad \quad \quad \quad \quad & 
$E_{21}/m$\quad \quad \quad \quad \quad \quad \\ 
\hline
0.0155522 & 0.9998785 & 0.9998791& 0.99987902866(7) & 0.99996976027(9) &
0.99996976506(8) \\
0.1425460 & 0.989458  & 0.989613 & 0.98960(4)       & 0.99742(0)       &
0.99745(2)       \\
0.2599358 & 0.96309   & 0.96364  & 0.9635(1)        & 0.9911(0)        &
0.9914(7)        \\
0.3566678 & 0.92578   & 0.92673  & 0.926(1)         & 0.982(3)         & 
0.983(8)         \\
0.4359255 & 0.88013   & 0.88139  & 0.88(0)          & 0.97(2)          & 
0.97(5)          \\
0.5000000 & 0.82758   & 0.82910  & 0.82(9)          & 0.96(1)          & 
0.96(7)          \\  
\hline
\end{tabular}
\end{table}

\begin{references}
\bibitem{mot} A detailed discussion about the interest of the study of the 
Herbst  equation can be found in Refs. [5], [6].
\end{references}
\end{document}